\documentclass[prl,twocolumn,amsfonts,showpacs]{revtex4-1}

\pdfoutput=1

\usepackage{epsfig}
\usepackage{psfrag}
\usepackage{amsmath}
\usepackage{amssymb}
\usepackage{color}
\usepackage{hyperref}
\usepackage{appendix}
\usepackage{bibunits}

\begin{document}
\author{Wei-Can Yang$^{1}$}
\author{Chuan-Yin Xia$^{2,1}$}
\author{Muneto Nitta$^{3}$}
\author{Hua-Bi Zeng$^{1}$}

\affiliation{$^1$ Center for Gravitation and Cosmology, College of Physical Science
and Technology, Yangzhou University, Yangzhou 225009, China}
\affiliation{$^2$ School of Science, Kunming University of Science and Technology, Kunming 650500, China}
\affiliation{$^3$ Department of Physics, and Research and Education Center for Natural Sciences, Keio University, Hiyoshi 4-1-1, Yokohama, Kanagawa 223-8521, Japan}

\title{Fractional and Integer Vortex Dynamics in Strongly Coupled Two-component Bose-Einstein Condensates from AdS/CFT Correspondence}

\begin{abstract}
In order to study the rotating strongly coupled Bose-Einstein condensations(BEC),
a holographic model defined in an AdS black hole that duals to a coupled two-component  condensations in
global $U(1)$ symmetry broken phase with intercomponent coupling $\eta$ and internal coherent coupling $\epsilon$ is proposed.
By solving the dynamics of the model, we study the process of formation and also the crossover from fractional to integer vortex phases.
With changing only $\eta$ from zero to a finite value,
fractional vortex lattices undergo a transition from hexagon to square lattice and finally to vortex sheets.
By continuing to turn on $\epsilon$, we find that two fractional vortices in different components constitute dimers, and when $\eta$ transcend a critical value, multi-dimer like hexamer or tetramer made up of two and three dimers appear. As $\epsilon$ keeps increasing, some dimers rotate to adjust themselves and then constitute the lattice of integer vortices.
Under an initial conditions similar to an spinor BEC vortices dynamics experiment, the appearance of disordered turbulence is found in the process of fractional vortex generation, which matches the experimental observation. While in the formation process of integer vortices, the appearance of  grooves is predicted.

\end{abstract}

\maketitle

\begin{bibunit}

The description of the nonequilibrium dynamics of strongly
interacting quantum many body systems remains one of the most challenging
problems in theoretical physics.
In such systems, e.g. in a rotating strongly coupled superfluid,
quantized vortices  play an important role and the the vortices formation
dynamics is always of great interest to study.
In weakly coupled systems, quantum vortices have been experimentally realized in
single component rotating Bose-Einstein condensates (BECs) of ultracold atomic gases
\cite{vortex-exp}.
The Gross-Pitaevskii (G-P) equations based on a weakly coupled
mean-field theory
perfectly simulate condensation behaviors and quantized vortices in BECs
at extremely low temperatures
 \cite{Dalfovo:1999zz,Pitaevskii:2003,Pethick:2008,Fetter}.
There are several
experiments which are described by weakly coupled theories,
such as superfluid He \cite{Tkachenko,Campbell}, ultracold atoms \cite{Madison,Abo-Shaeer,Lin} and Fermi superfluids \cite{Zwierlein}.
However, we need to go beyond weakly coupled theories to describe
a strongly coupled BECs at a finite temperature, which is still a challenging problem \cite{Allan}.
For decades, the anti-de Sitter(AdS)/conformal field theory(CFT) correspondence \cite{Maldacena,Witten,Gubser,Aharony:1999ti}
or holography has been proved to be
a unique ``first principle" method to study  equilibrium strongly coupled quantum systems at a finite temperature or zero temperature,
by solving their dual weakly coupled gravity theory \cite{Zaanen:2015oix,Ammon2015}. Also,  the holographic method has successfully  simulated the dynamics of a strongly coupled  system far from equilibrium \cite{Liu:2018crr,Liu,Sonner:2014tca,Bhaseen,Chesler:2014gya,Zeng:2019yhi},
then it is a perfect method to study the vortex formation dynamics and equilibrium state in a rotating single component strongly coupled BECs \cite{Xia,Tianyu} and also in the vortex for superconductor in a magnetic field \cite{Gianni,Kengo,Aristomenis}.

\begin{figure}[t]
\centering
\includegraphics[trim=1.95cm 11.5cm 1cm 9.8cm, clip=true, scale=0.5, angle=0]{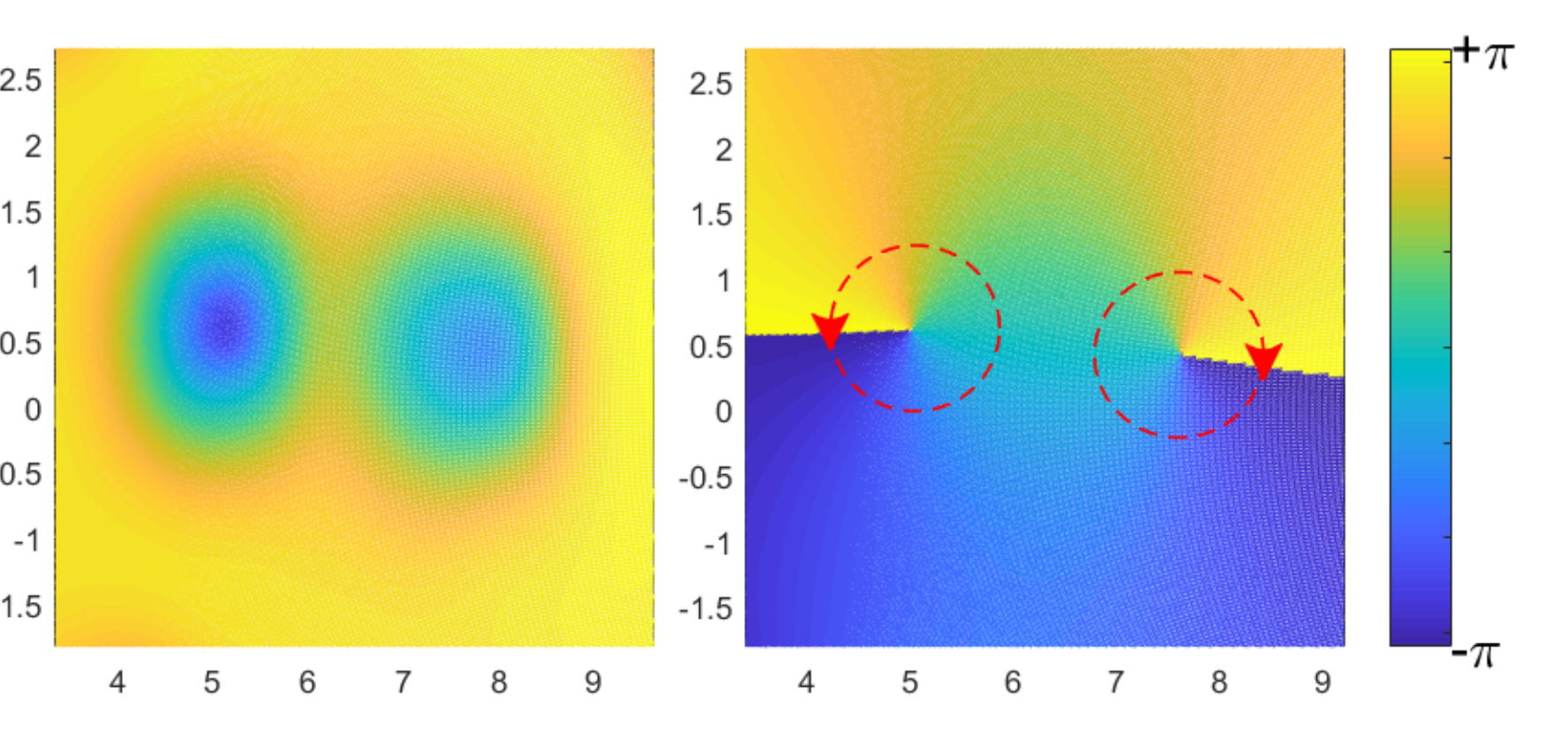}

\caption{(left): A vortex pair in a vortex lattice with $\eta=3$ and $\epsilon=0.005$. (right): The phase difference between the two components: $\phi(r)=\theta_1(r)-\theta_2(r)$ with the form that $\langle O_1\rangle=|\langle O_1\rangle|e^{i\theta_1}$ and  $\langle O_2\rangle=|\langle O_2\rangle|e^{i\theta_2}$. }\label{fig1}
\end{figure}

Rather than a single component BEC,
the two-component BECs
of $^{87}$Rb
with the internal coherent (Rabi or Josephson) coupling were also experimetally realized by the JILA group \cite{Hall,coherent}.
With the development of study of vortices in ultracold atomic gases,
many attentions have been paid to phenomena of vortices in multi-component condensations
\cite{Son:2001td, Mueller,Ueda,Kasamatsu,Tsubota,Kasamatsu-2,
Galteland,Kobayashi:2018ezm,
Aftalion,Kuopanportti,Kasamatsu-4,Eto:2011wp,Eto-2,Eto:2013spa,Nitta,Cipriani:2013wia,Dantas:2015fka,Tylutki:2016mgy,Orlova:2016,Calderaro:2017,Eto:2017rfr,Kasamatsu:2015cia,Uranga,
Eto:2019uhe,Shinn:2018zde}.
In contrast to single-component BECs where only hexagonal Abrikosov lattice can form, a large number of lattice structures have been found in the two-component BECs.
Such lattice structures are also expected in unconventional superconductors with multiple gap functions.
When the intercomponent coupling is gradually increased, one observe transformations from the Abrikosov's hexagonal lattices to square lattices, stripe lattices and sheet lattices \cite{Ueda,Kasamatsu}. Experimentally, a square lattice in two component spinor BEC has been found \cite{Schweikhard} and, more recently, a special lattice structure has been found in the mixture of  Bose-Fermi superfluid with different masses \cite{Yao}.
One of alluring features in multi-component condensations is that
quantized circulations for superfluids
\cite{Son:2001td, Mueller,Ueda,Kasamatsu,Tsubota,Kasamatsu-2,
Galteland,Kobayashi:2018ezm,
Aftalion,Kuopanportti,Kasamatsu-4,Eto:2011wp,Eto-2,Eto:2013spa,Nitta,Cipriani:2013wia,Dantas:2015fka,Tylutki:2016mgy,Orlova:2016,Calderaro:2017,Eto:2017rfr,Kasamatsu:2015cia,Uranga,Eto:2019uhe}
or fluxes for superconductors
 \cite{Babaev,Goryo,fractional-exp,Tanaka}
are not integer-valued anymore but are rational or fractional.
In the order-parameter field, an integer quantized vortex has
a winding $\phi\rightarrow\phi+2\pi$ while a fractional quantized vortex does a winding $\phi\rightarrow\phi+\pi$
when the order-parameter phase $\phi$ passes through a path around the vortex core \cite{Autti}. However, when the internal coherent coupling $\epsilon$ is added
in two-component BECs of $^{87}$Rb experiments \cite{Hall,coherent},
 two fractional vortices in different components form a vortex pair bound by a domain wall, as shown in FIG.~\ref{fig1}, and richer lattice structures would appear as predicted theoritically \cite{Tsubota,Nitta}. Finally, the two vortices overlap and form a integer vortex when $\epsilon$ exceeds a critical value \cite{Nitta}.



\begin{figure}[t]
\centering
\includegraphics[trim=1.4cm 12cm 1cm 9.1cm, clip=true, scale=0.53, angle=0]{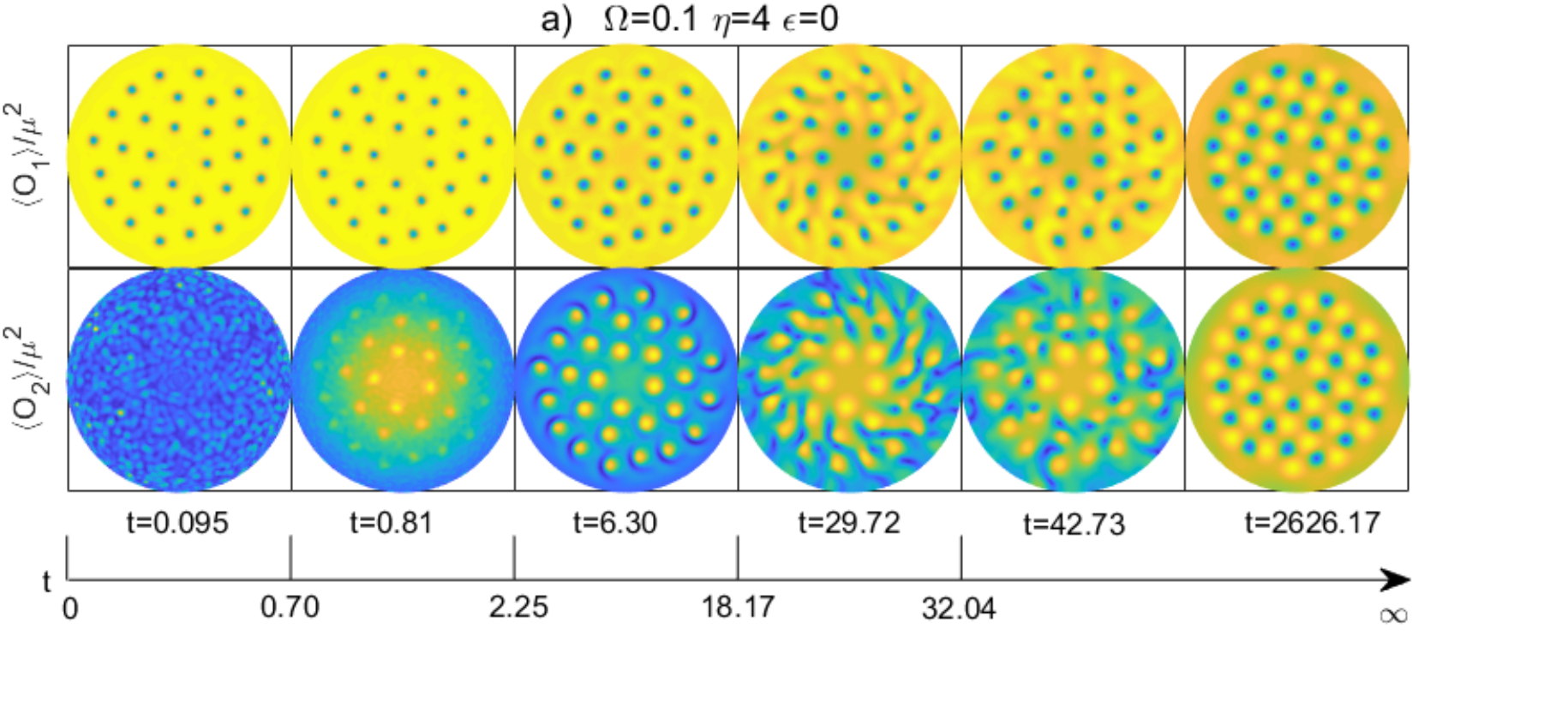}
\includegraphics[trim=1.4cm 12cm 1cm 9.1cm, clip=true, scale=0.53, angle=0]{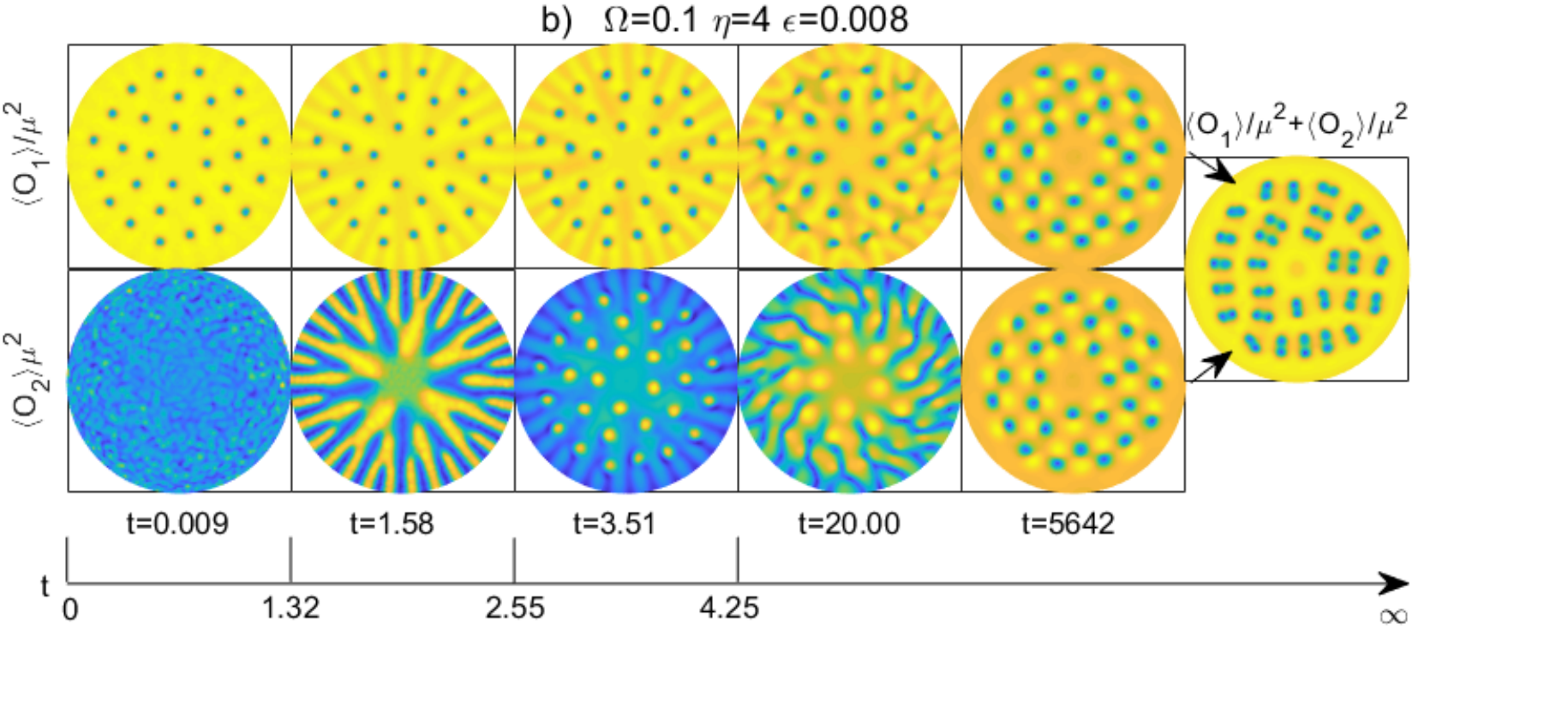}
\includegraphics[trim=1.4cm 12cm 1cm 9.1cm, clip=true, scale=0.53, angle=0]{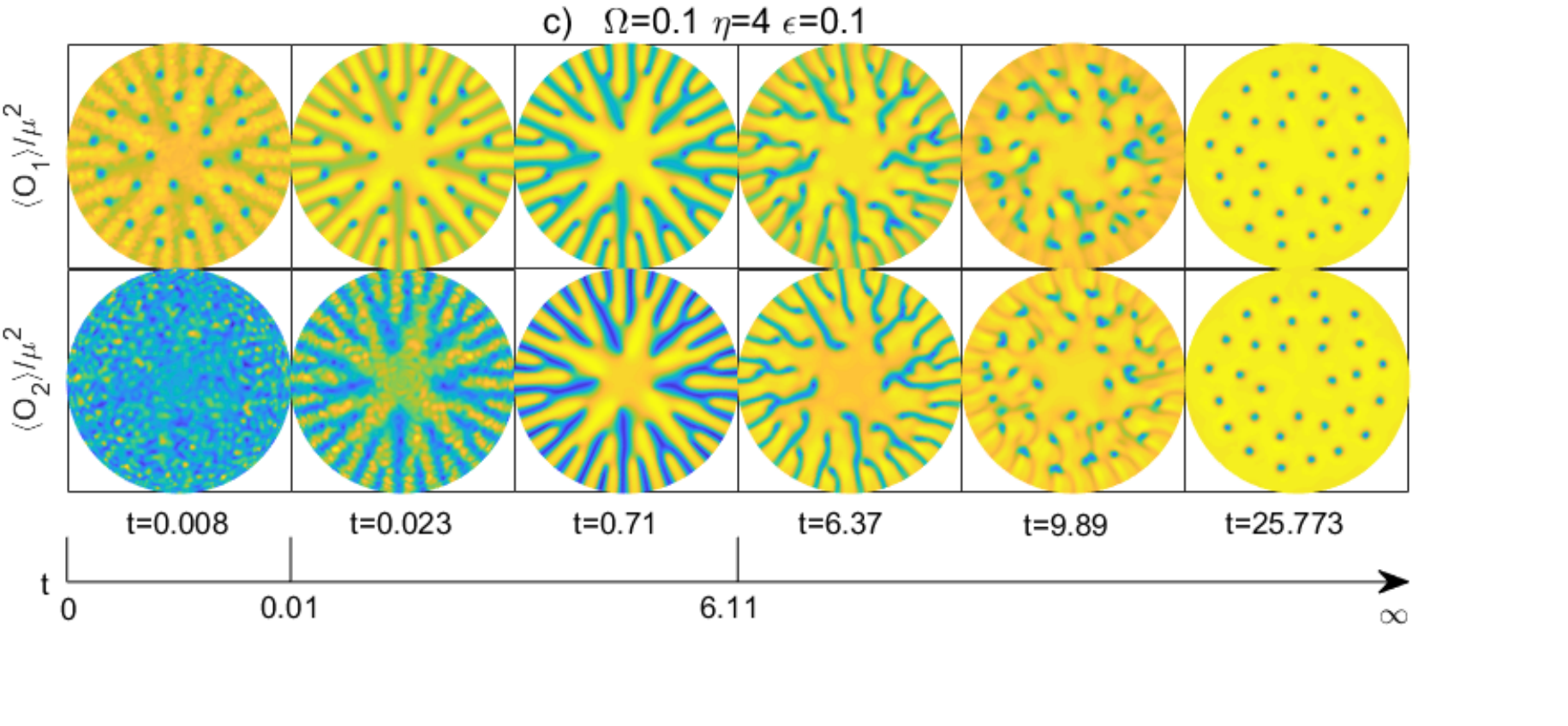}

\caption{The dynamics for three different vortex phases formation.
 (a) fractional vortices without internal coherent coupling, (b) dimers,
 (c) integer vortices.
.}\label{fig2}
\end{figure}

\begin{figure*}[t]
\centering
\includegraphics[trim=0cm 5cm 0cm 5cm, clip=true, scale=0.9, angle=0]{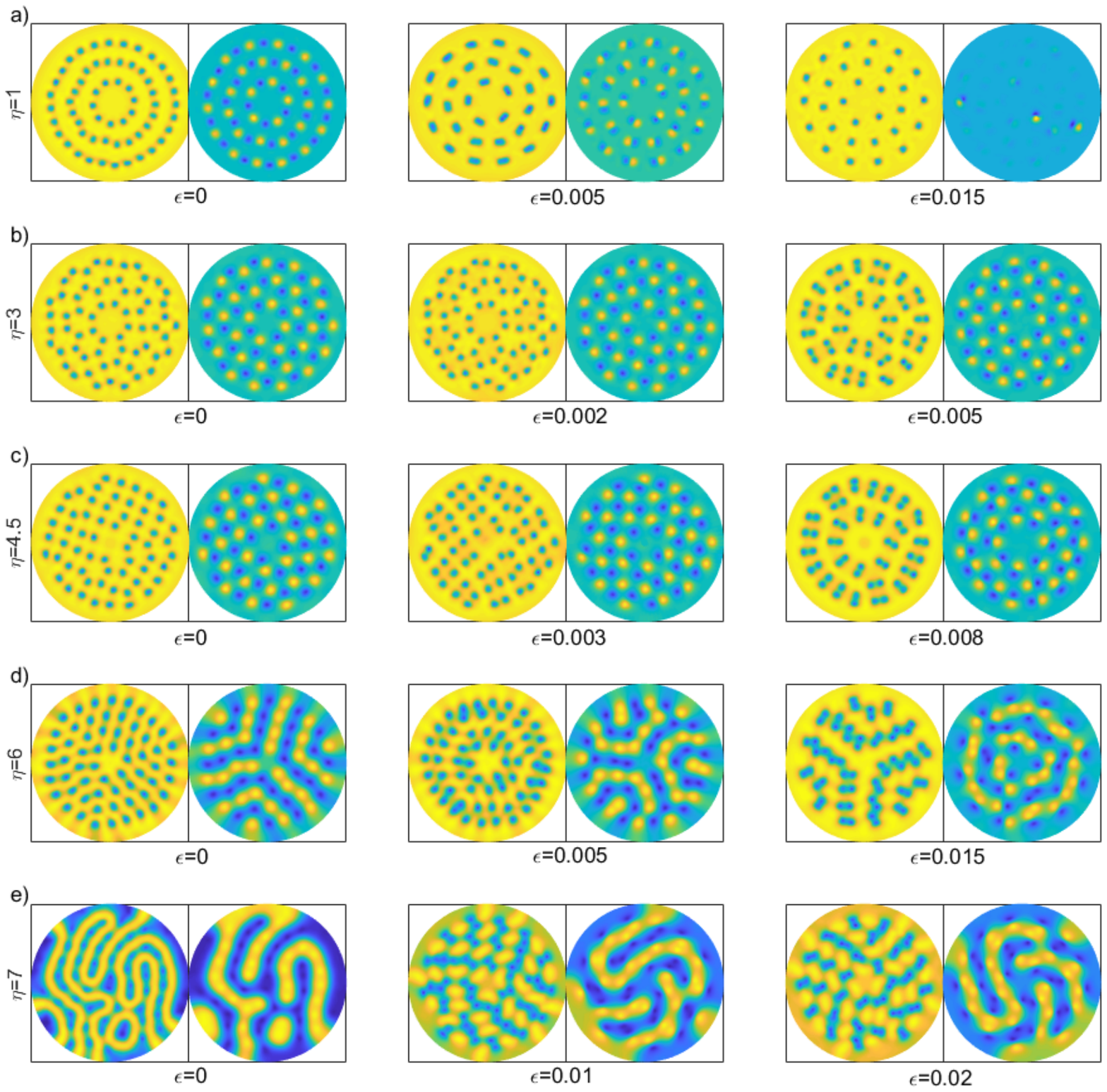}

\caption{The final equilibrium fractional and integer vortices by changing different $\epsilon$ with fixed $\eta$ after the formation of two-component vortex lattices.
In all subfigures the left is a plot of the density profile of the conedensate $n = |\langle O_1\rangle|+|\langle O_2\rangle|$, the right panel is a figure of $|\langle O_1\rangle|-|\langle O_2\rangle|$ that displays $\langle O_1\rangle$ and $\langle O_2\rangle$ vortices, respectively, where blue parts represent vortices of $\langle O_1\rangle$ while yellow parts represent vortices of $\langle O_2\rangle$. The values of the coupling coefficient $\eta$ and $\epsilon$ are shown in the left and bottom, respectively, of each case. }\label{fig3}
\end{figure*}

In this paper, we study the dynamic formation process of fractional and integer vortices in strongly coupled BECs made of two components and the dynamic crossover from fractional to integer vortices in strongly interacting systems, in the frame work of the AdS/CFT correspondence. We first use the time-dependent evolution of a single component to obtain a stable hexagonal Abrikosov lattice, then switch on the second component with the same chemical potential.
We thus find a bizarre turbulence phenomenon
and grooves
in the coupling ranges of fractional vortices and integer vortices, respectively.
These were not seen in the previous theoretical analysis \cite{Kanai},
but are similar to the experiment \cite{Schweikhard}. Two components behave independently like a single component BECs when the two couplings constants between the two components are vanishing, while once we turn on the coupling $\eta$, there will appear a rich lattice structure, as shown in
FIG.~\ref{fig2}. Subsequently turning on $\epsilon$, the vortices in different components will form dimers. In the dynamic evolution, we see the autonomous arrangement of the dimers in  the minimum free energy state, which is confirmed  in FIG.~\ref{fig5} shown in \cite{details}, finally forming a lattice structure of integer vortices,
as shown in FIG.~\ref{fig3}.


The holographic model is constructed by a dual of an asymptotically AdS spacetime in three spatial dimensions with the quantum conformal field (AdS$_{3+1}$/CFT$_{2+1}$). As found in Ref.~\cite{Hartnoll}, to construct a quantum field theory on the AdS boundary of two spatial dimensions,  a complex scalar  $\Psi(x)$, carrying charge $q$ under a global $U(1)$ symmetry is proposed to live in the AdS$_4$ black hole. Let $j_\mu(x)$ denote the conserved current operator of this global $U(1)$ symmetry. To induce a superfluid condensate for $\Psi$, we will turn on a chemical potential $\mu$ for the $U(1)$ charge. For sufficiently large $\mu$, we
expect $\Psi$ to develop a nonzero expectation value on the boundary when the temperature falls below
a critical temperature $T_c$, spontaneously breaking the global $U(1)$ symmetry and driving the
system into a superfluid phase. To simulate a two component BECs with the same chemical potentials, there are two complex bulk charged scalar fields $\Psi_1$ and $\Psi_2$, with charge $q$ and the mass $m_j$, separately coupled with two $U(1)$ gauge fields $A_{\mu(1,2)}$,
thus yeilding two chemical potentials $\mu_j$.
We then have the bulk action
\begin{equation}
S=\int  d^4x \sqrt{-g}\Big[-\sum_{j=1}^2(\frac{1}{4}F_j^2+ |D_j\Psi_j|^2+m_j^2|\Psi_j|^2)+ V
\Big]
\label{model}
\end{equation}
The inter-component coupling potential between the two charged scalar fields takes the form
\begin{equation}
V(\Psi_1,\Psi_2)= \epsilon(\Psi_1\Psi_2^*+\Psi_1^*\Psi_2)+\eta |\Psi_1|^2|\Psi_2|^2,
\label{potential}
\end{equation}
where $F_{ j\mu\nu}=\partial_\mu A_{ j\nu}-\partial_\nu A_{ j\mu}$, $D_{j\mu }=\partial_\mu-iqA_{j\mu}$ with $q$ the charge. Here,
$\eta$ is the intercomponent coupling coefficient and $\epsilon$ is the internal coherent (Rabi) coupling coefficient.
Notice that the two components are decoupled when $\eta=\epsilon=0$, the equations of motion and the
numerical scheme can be found in \cite{details}.
 This situation is different from that of the model proposed in
 Ref.~\cite{Wen:2013ufa},
where the two components are always coupled even when both the coupling constants vanish,
the vortex diagram of the model was studied and no fractional vortices were found \cite{Yang}.

In FIG.~\ref{fig2}. we show the evolution processes of the formation of vortex lattices from the holographic model Eq.~\ref{model},
when $\mu_1=\mu_2>\mu_c$.
We set the rotation speed to $0.1$ for all dynamic processes.  We first turn off the $\langle O_2\rangle$ component by the vanishing  condition
for $\Psi_2$ in the dynamical process, which turns the kinetic equation into a single component equation like that in Ref.~\cite{Hartnoll}. We can see that the $\langle O_1\rangle$ component forms a hexagonal lattice like the first image of $|\langle O_1\rangle|$ in FIG.~\ref{fig2} (a). Then we give the $\langle O_2\rangle$ component a slight perturbation,
 turn off the vanishing  condition for $\Psi_2$ and also increase the $\eta$. Different evolution processes are illustrated in different $\epsilon$ values, such as the three subfigures in FIG.~\ref{fig2}. We take the $\eta=4$, and take $\epsilon=0$, $\epsilon=0.008$ and $\epsilon=0.1$ in subfigures (a), (b), (c), respectively, to represent the fractional vortex range of the non-attractive interaction, dimer range and integer vortex range, respectively.

In the range $0<t<0.7$ of fractional vortices,
the $\langle O_2\rangle$ has no change. On the other hand, when $0.7<t<2.25 $, a condensation appears at the corresponding position of $\langle O_1\rangle$ vortices, which reflects the separation of the two components. When $2.25<t<6.3$, semicircular ripples appear around the condensation, and the direction of the ripples are opposite to that of rotation, which can be seen visually in the movie in the
Supplemental Material.
As evolution continues, these ripples rotate quickly and gradually spread out, forming a disorderly turbulence.  This turbulence extends to the center of rotation and eventually vortices form directly in the turbulence when $t=32.04$. Both the turbulence and the vortices forming in it are separated from the vortices of $\langle O_1\rangle$. Over time, vortices of the two components form a square lattice in the end.

In the range of integer vortices, different phenomenon happens. In a very short time of about $t=0.01$, the two components respond. Exactly the same grooves form in the two components when $0.01<t<6.11$. These grooves gradually shrink until they form a complete superimposed integer vortex.

When $\epsilon$ is in the range of dimers, that is, the range between non-attractive interactions and integer vortices, we find that the vortex evolution process is like a combination of the two previous processes. The grooves and turbulence appear in the front and back stages, respectively. We infer that this is because in the initial stage, $\langle O_2\rangle$ has fewer components, which is difficult to condense at the position of the $\langle O_1\rangle$ vortices, and so the separation cannot be shown. This indicates that the turbulence is caused by the separation interaction.

\begin{figure}[t]
\centering
\includegraphics[trim=2.2cm 9.9cm 1cm 9.6cm, clip=true, scale=0.55, angle=0]{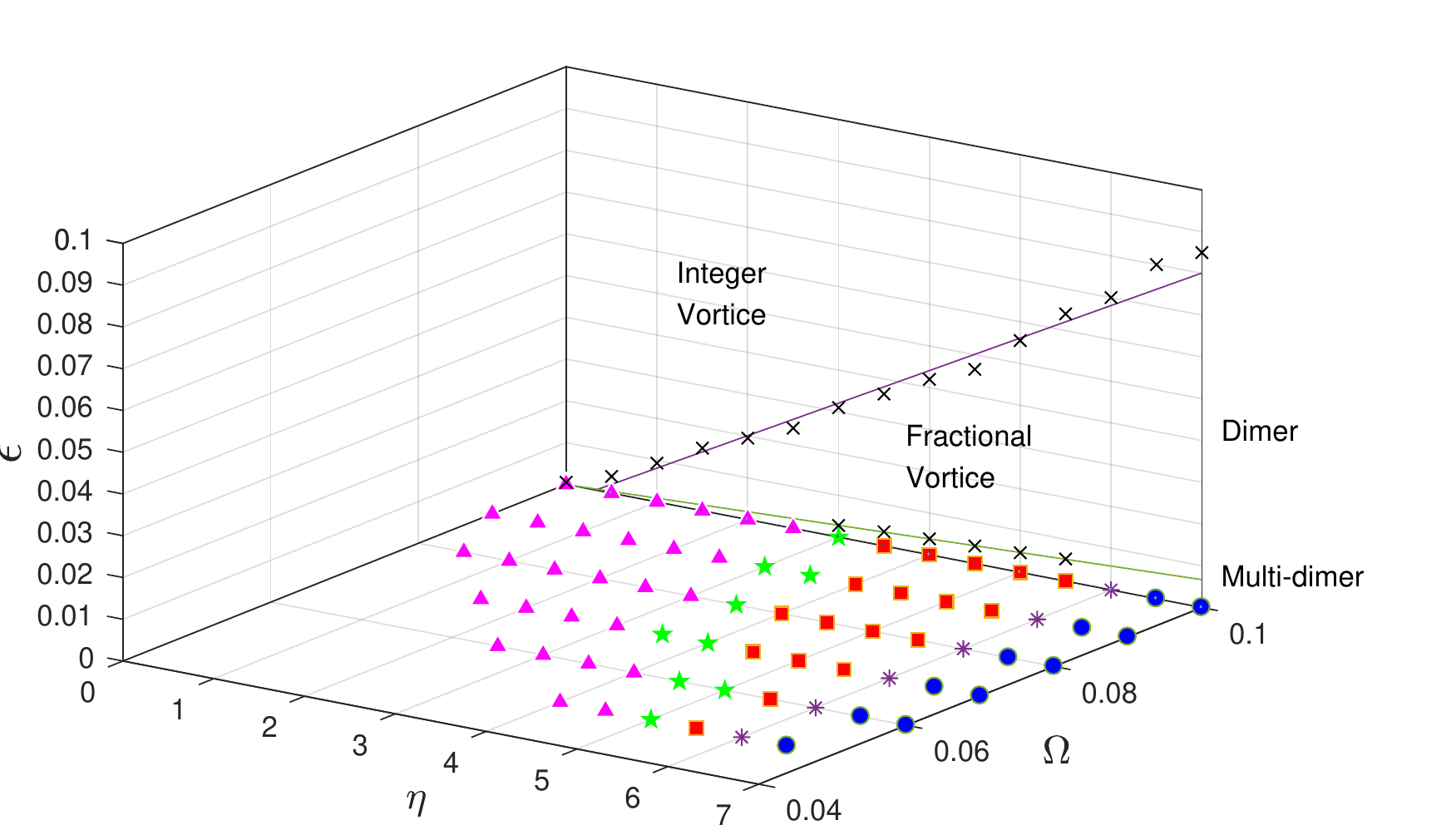}
\caption{(Bottom): the $\Omega-\eta$ phase diagram of the vortex states, where
$\triangle$,
$\star$,
$\square$,
$\times$,
 $\circ$, denote
triangular lattices,
triangular tetragonal cross lattices,
square lattices,
stripe vortices and
vortex sheet lattices, respectively.
(Side): the $\epsilon-\eta$ phase diagram with the boundary between fractional and integer vortices: $\epsilon=0.012\eta-0.0039$ and the boundary between dimers and multi-dimers :$\epsilon=0.00096\eta-0.000047 $ }\label{fig4}
\end{figure}

Next we investigate the dynamic evolution from fractional to integral vortices by gradually increasing $\epsilon$, as shown in FIG.~\ref{fig3}.
Subfigures (a,b), (c), (d) and (e) represent hexagonal, square, stripe and sheets lattices, respectively.
Because the dividing line of multi-dimer/dimer and fractional/integer vortex lattices are approximatively linear between $\eta$ and $\epsilon$,
it is convenient to  use the ratio relation between $\eta$ and $\epsilon$ to discuss the process. When $0<\epsilon/\eta<0.00096$ and $3<\eta<5.5$, we find that multi-dimer bound states appear. When $\eta$ takes values corresponding to hexagonal lattices, hexamers composed of three dimers form, as shown in middle part of FIG.~\ref{fig3}(b). If $\eta$ is
within the square lattice range, each bound state is composed of four vortices making up tetramers, as shown in the middle part of FIG.~\ref{fig3}(c). However, different from the G-P equation, there are no multi-dimers when $\eta<3$ in our model. On the other hand, the phase separation for $\eta>5.5$ always prevents the formation of multi-dimer bound states for extremely low value of $\epsilon/\eta$.
When $0.00096<\epsilon/\eta<0.012$, multi-dimers split into dimers. We find that dimers rotate to adjust themselves to meet the minimum energy requirements, which can be easily observed by comparing the middle and right figure in FIG.~\ref{fig3}(b), (c), (d), (e). This rotation behavior also results in rich rotational symmetric lattice structures. While for $5.5<\eta$, the free energy is minimized when vortices of the same component are close to each other, forming stripe or sheet lattice, so for $\epsilon$ in the range of the non-integer vortices, there is always a phase separation. As $\epsilon$ continues to increase to the condition  $\epsilon/\eta>0.012$, vortices in different components overlap to form an integer vortex lattice, as shown in the right figure of FIG.~\ref{fig3}(a), where $|\langle O_1\rangle|-|\langle O_2\rangle|$ is almost zero.

By organizing the results, we obtain the phase diagram of fractional and integer vortex lattices with $0.05<\Omega<0.1$, $0<\eta<7$ and $0<\epsilon<0.1$. The results are illustrated in FIG.~\ref{fig4}. The bottom phase diagram of $\Omega-\eta$ is similar as that in the G-P equations \cite{Ueda} and a simplified holographic model \cite{Yang}. When we take $\Omega=0.1$ as specified above, by increasing $\eta$ from 0, we can see hexagonal lattices ($0<\eta<3$), square lattices ($3<\eta<5.5$), stripe lattices ($5.5<\eta<6.5$) and eventually sheet lattices ($6.5<\eta$). When $\epsilon$ continues to increase, two vortices in each of the two components couple to form a dimer, and when $\eta$ is in the range of $(3-5.5)$, three or two dimers would further couple to form multi-dimers, such as hexamers and tetramers. By fitting the boundary between dimer and multi-dimer and between fractional and integer vortices, we can see two linear relationship between $\eta$ and $\epsilon$, similar to the results from the G-P equations \cite{Nitta}.

In summary, with the power of the AdS/CFT correspondence, we have been able to  access to the strongly coupled dynamics of a two component BEC by solving the classical gravitational
equations in the bulk AdS space. A unique phenomenon of turbulence and grooves during the generation of fractional and integer vortex lattices have been observed. By comparing the generation process of different parameters, we have found that the turbulence is related to the separation interaction. In addition, we have studied the evolution process from the fractional vortex lattice to the integer vortex lattice with the increase of the internal coherent coupling $\eta$. We have found many differences with the results from the G-P equation, including the criticality of the multi-dimer, the rotation of dimers and rich lattice structures with slight differences.

\emph{Acknowledgements}.---
 This work is supported by the National Natural
Science Foundation of China (under Grant No. 11675140).
This work of MN was supported in part by the
Ministry of Education, Culture,
Sports, Science (MEXT)-Supported Program for the Strategic Research
Foundation at Private Universities ``Topological  Science''
(Grant  No.~S1511006),
Japan Society for the Promotion of Science (JSPS) KAKENHI
(Grant Numbers 16H03984 and 18H01217)
and by
a Grant-in-Aid for Scientific Research on Innovative Areas
``Topological Materials Science''
(KAKENHI Grant No.~15H05855) from MEXT of Japan.

\end{bibunit}

\begin{bibunit}

\begin{appendices}

~
\newpage
\section{SUPPLEMENTAL MATERIAL}

\subsection{Equation of motion and numerical scheme}

The model proposed in Eq. \ref{model}  is living in a black hole whose metric in Eddington coordinate can be written as
\begin{equation}
ds^2=\frac{\ell^2}{z^2}\left(-f(z)dt^2-2dtdz + dr^2+ r^2d\theta^2\right)
\tag{1}
\label{1}
\end{equation}
in which $\ell$ is the AdS radius, $z$ is the AdS radial coordinate of the bulk
and $f(z)=1-(z/z_h)^3$.  Thus, $z=0$ is the AdS boundary while $z=z_h$ is the horizon; $r$ and $\theta$ are the radial and angular coordinates. The Hawking temperature is $T=3/(4\pi z_h)$. Without loss of generality we can rescale $q=\ell = z_h = 1$.
The probe limit is adopted in the paper by assuming that the matter fields do not affect the gravitational fields,
as long as one neglects the effect of the superfluid condensate on the normal component of
the fluid \cite{Liu}. This is the same as the G-P equations where the normal component is not considered in the model.
The equations of motion are simply
\begin{equation}
d_\nu F^{\mu\nu}_j = J^\mu_j\tag{2}
\end{equation}
\begin{equation}
(-D_j^2+m_j^2+\epsilon+\eta|\Psi_k|^2)\Psi_j=0\tag{3}
\end{equation}
where $j=1,k=2$ or $j=2,k=1$.

Under the standard holographic dictionary, the bulk scalar field $\Psi(r,\theta,z)$ can be mapped to a scalar operator $\Psi^i(r,\theta)$ of a quantum  field theory at the boundary and $A_\mu(r,\theta,z)$ can be mapped to the conserved current $J^\mu(r,\theta)$. The axial gauge $A_z=0$ is adopted as in Ref. \cite{Liu}.  Near the boundary, the general solutions take the asymptotic form as $A_{j\nu}=a_{j\nu}+\mathcal{O}(z)$, $\partial_z A_{j\nu}= b_{j\nu}+\mathcal{O}(z)$,
($j=1,2$). The coefficients $a_{jr,j\theta}$ can be regarded as the superfluid velocity along $r, \theta$ directions while $b_{jr,j\theta}$ as the conjugate currents \cite{Montull}.
$a_{jt}$ corresponds to the charge density $\rho$ of the field theory, while $b_t$ is the chemical potential
$\mu$ of  which is inversely proportional  to the dimensionless temperature $T \propto 1/\mu$. The $U(1)$ symmetry is spontaneously broken below the critical temperature, $T_c$, so as to increase the $\mu$ to a critical $\mu_c$, then the BECs form. In the broken phase, the field $\Psi$ has an non-zero solution whose expansion coefficient on the boundary is the expectation value of the dual operator in the field theory. By choosing that $m_1^2=m_2^2=-2$ with $\Delta=2$, one can obtain $\partial_z \Psi_j=\Psi_j^1+\mathcal{O}(z)$, $\partial_{zz}\Psi_j =2\Psi_j^2 +\mathcal{O}(z)$, where $\Psi_j^1$ is a source term which is always set to be zero in the spontaneous symmetry broken phase, $\Psi_j^2$ is the vacuum expectation value $\langle O_j\rangle$ of the dual scalar operator, which is non-vanishing in the broken phase.

The equations of motion (EoMs) are solved numerically by the Chebyshev spectral method
in the $z,r$ direction, while the  Fourier decomposition is adopted in the $\theta$ direction. The radius of the boundary disk is set as $r=R=20$. The Neumann boundary conditions are adopted both at $r=R$ and $r=0$, $\partial_r h_i=0$ where $h_i$ represents all the fields except $a_{j\theta}$.
The rotation is introduced by imposing the angular boundary condition as $a_{j\theta}=\Omega r^2$ \cite{Domenech},where $\Omega$ is the constant angular velocity of the disk.

From the numerical simulation of the non-rotating state with the Newton-Raphson method we found that $\mu_c \sim 4.07$ when $\epsilon=\eta=0$,
so the dimensionless critical temperature is $T_c^0 = \frac{3}{4\pi \mu_c(0,0)}=0.0587$.
With the fixed $\mu=6$, we obtain the uniform solution in the non-rotating state with the Newton-Raphson method, which can be served as the initial value of dynamic simulation. Begin with this initial uniform solution, the time evolution is simulated by the fourth order Runge-Kutta method. By setting the $\eta$ , $\epsilon$ and $\Psi_2$ to zero, we obtain a single component EoMs, which end up with a hexagonal lattice. Then, given a finite value of $\eta$ and $\epsilon$ and a perturbation of $\Psi_2$, one can obtain the evolution of two-component vortex lattices similar
to the experiment \cite{Schweikhard}.

\subsection{Free Energy}

At a fixed temperature, the free energy $F$ can be computed from the renormalized on-shell action $S_{ren}$ with two parts $S_{o.s.}$ and $S_{c.t.}$. $F=TS_{ren}=T(S_{o.s.}+S_{c.t.})$, in which $S_{o.s.}$ is the bare on-shell action obtained by subtracting the equation of motions from Eq.\eqref{model} while $S_{c.t.}=-\sum_{j=1}^2\int dtdrd\theta\sqrt{-\gamma}\Psi_j\Psi_j^\star|_{z=0}$ is the term to remove the divergence near the boundary $z=0$, where $\gamma$ is the determinant of the reduced metric. Then, one can obtained the form of the renormalized on-shell action that

\begin{figure}[h]
\centering
\includegraphics[trim=3.3cm 11cm 1cm 11cm, clip=true, scale=0.6, angle=0]{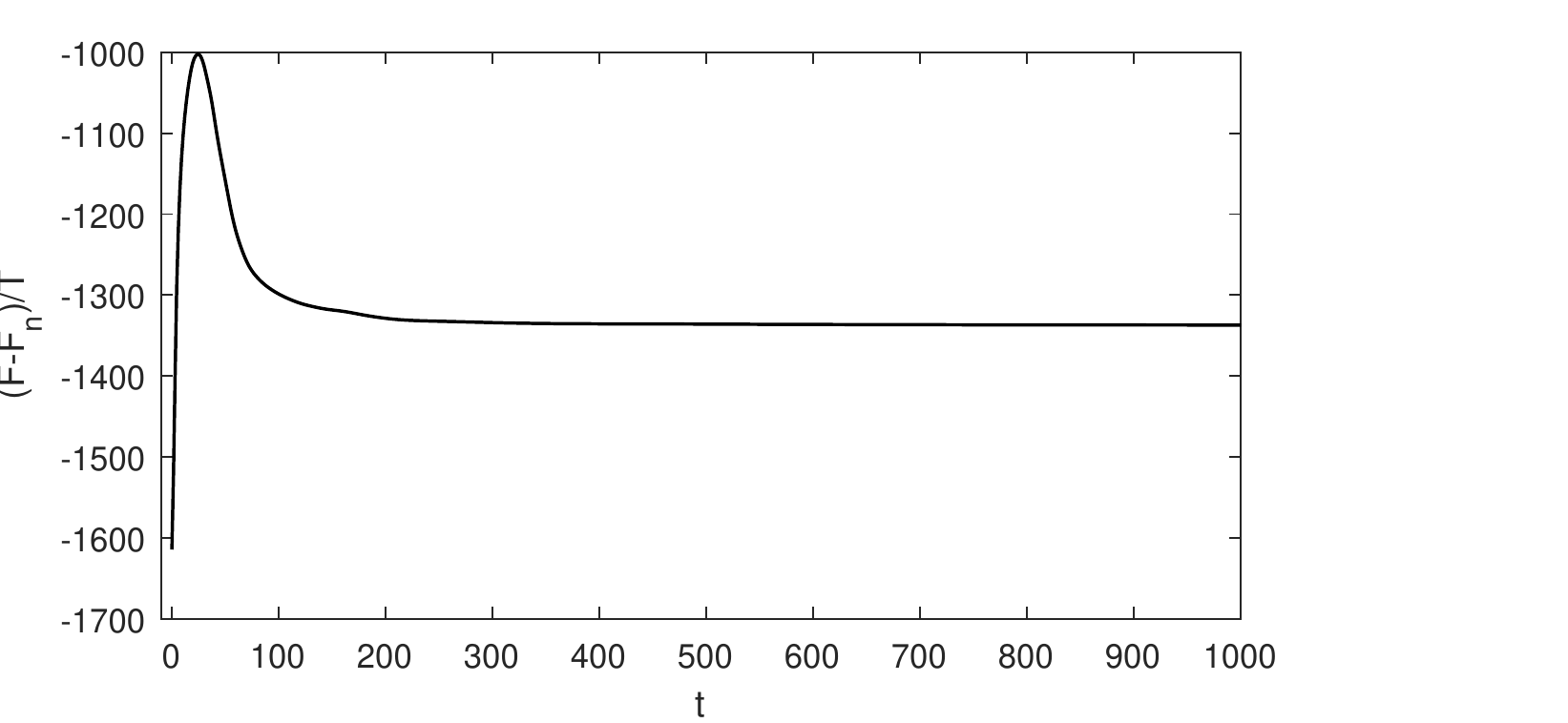}
\caption{The time evolution of rescaled free energy with $\Omega=0.1$, $\eta=4$ and $\epsilon=0$.}\label{fig5}
\end{figure}

\begin{widetext}
\begin{equation}
\begin{split}
S_{ren}=\frac{1}{2}\sum_{j=1}^2\left\{-\int dtdzd\theta[\frac{1}{r}A_{\theta j}\partial_rA_{\theta j}]|_{r=R}+\int dtdrd\theta[r(-a_{tj}b_{tj}+\frac{1}{r^2}a_{\theta j} b_{\theta j}+\Psi_j^{1\star}\Psi_j^2+\Psi_j^{2\star}\Psi_j^1)\right\}\\ +\frac{iq}{2}\sum_{j=1}^2\int dtdzdrd\theta[\frac{r}{z^4}A_{\mu j}(\Psi_j^\star\partial^\mu\Psi_j-\Psi_j\partial^\mu\Psi_j^\star)-2iqA^\mu_j|\Psi_j|^2] ]\\
+\int dtdzdrd\theta [\epsilon(\Psi_1\Psi_2^\star+\Psi_1^\star\Psi_2+|\Psi_1|^2+|\Psi_2|^2)+3\eta|\Psi_1|^2|\Psi_2|^2
] \end{split}
\end{equation}
\end{widetext}
In the Fig.\ref{fig5} we calculated the free energy $(F-F_n)/T$ of a dynamic process with parameter that $\Omega=0.1$, $\eta=4$ and $\epsilon=0$, in which $F_n$ is the free energy in the normal state, i.e., $\Psi_j=0$.. It can be seen that after $t=500$, the free energy reaches its lowest point and doesn't fluctuate any more. Then, we can say that after this time, the system reaches equilibrium.

\end{appendices}
\end{bibunit}

\end{document}